\documentclass[aps,prb,twocolumn]{revtex4}

\usepackage{amsmath}    
\usepackage{graphicx}   

\begin{document}

\title{Computing Accurate Age and Distance Factors in Cosmology}

\author{Jodi L. Christiansen}
 \affiliation{California Polytechnic State University, San Luis Obispo, CA 93407}
 \email{jlchrist@calpoly.edu}   
\author{Andrew Siver}
 \affiliation{California Polytechnic State University, San Luis Obispo, CA 93407}
\date{\today}

\begin{abstract}
As the universe expands astronomical observables such as brightness and angular size on the sky change in ways that differ from our simple Cartesian expectation.  We show how observed quantities depend on the expansion of space and demonstrate how to calculate such quantities using the Friedmann equations. The general solution to the Friedmann equations requires a numerical solution which is easily coded in any computing language (including EXCEL).  We use these numerical calculations in four student projects that help to build their understanding of high-redshift phenomena and cosmology. Instructions for these projects are available as supplementary materials.
\end{abstract}

\maketitle

\section{Introduction}
In the past decade the field of observational cosmology has made significant progress.  Recent observations have dramatically improved density estimations for the primary components of the universe.  Two of these components, dark matter and dark energy, have so far eluded direct detection although a variety of indirect measurements clearly point to their existence.  Something as simple as the dynamics of a rotating galaxy implies that galaxies contain large amounts of unseen mass---the so-called cold dark matter (CDM).  Also, the observed reddening of distant supernovae implies that the expansion of the universe is accelerating, powered by Einstein's famous cosmological constant ($\Lambda$) and attributed to some not-yet-understood dark energy.  

A surprisingly simple model of the Universe that uses energy conservation to relate the spatial expansion to the energy density can account for the increasingly detailed and accurate observations now becoming available.  This model, embodied in the Friedmann equations, has resulted in a widely accepted Standard Model of Cosmology.~\cite{standardModel}  In this paper we show how the Friedmann equations relate to astronomical observables such as brightness and angular size on the sky.~\cite{ryden, peebles, hogg}  Moreover, we show how students can use these relations to compute accurate distances and ages just as professional astronomers do.

Although no analytic solution exists for the general Friedmann equations, simplifying assumptions can be used to solve a number of interesting special cases.  Taylor expansions can also be used to compute approximate distances and ages, but for our universe such expansions are only accurate to a distance where the redshift ($z$) is around 0.1.  A redshift of 0.1 corresponds to light emitted when the universe was already 90\% of its current size.  To study events occurring at earlier times, when the universe was a tiny fraction of it's current size, we need accurate calculations out to redshifts of 1000 or more. 

To go beyond analytic solutions requires a computational approach, which can be both instructive and rewarding for students.  Besides giving students practice using numerical techniques, these computations allow them to accurately compute ages and distances back to the time of the Big Bang.  Reproducing the latest published age and size of the visible universe builds student confidence and removes much of the mystery of where these numbers come from.  Students can also compute the physical size of high-redshift objects (in their rest frame) and compare such objects to our own galaxy or other nearby galaxies.  It is a powerful learning experience for a student to deduce that seven billion years ago gravitational attraction formed structures similar to the ones visible in our night sky.

In this paper we give a brief description of the quantities that astronomers measure and how such quantities relate to the spatial expansion of the universe.  We then present the solution to the Friedmann equations and the numerical techniques used to solve them.  Finally, we present four important projects that make use of these computational solutions to help students understand high-redshift phenomena. 

\section{Astronomical Observables}
Telescopes point at astronomical sources to measure, among other things, their light intensity, spectra, and angular extent on the sky.  At large distances these observables depend on the geometry and expansion rate of the universe.  In fact, the expansion leads directly to an observed reddening of distant objects---the so-called redshift.  We define this redshift in terms of the wavelength of observed light.  If a distant galaxy emits light of wavelength $\lambda_e$, and we observe (redshifted) light of wavelength $\lambda_{ob}$, the redshift is defined by
\begin{equation}
z = \frac{\lambda_{ob} - \lambda_e}{\lambda_e}.
\label{eq:redshift}
\end{equation}

Spectral lines of hydrogen, helium, and a number of other elements are routinely measured in undergraduate laboratories.  In astronomy, a spectrometer can be used to measure the observed wavelength of astronomical objects with strong spectral lines.  Because we already know the emitted wavelengths from our laboratory studies we can easily determine the redshift.  This redshift can result from a Doppler shift due to the velocity of the astronomical source or from the expansion of the universe.  In general, redshifts are a combination of the two.  Galaxy cluster velocities~\cite{peculiar} almost never exceed $\sim$\,1\,000\,km/s, which corresponds to a Doppler redshift of 0.003.  Redshifts of cosmological origin are often significantly higher than this.  Beyond a redshift of $z\sim 0.03$, the expansion of the universe generally dominates and the Doppler shift can be neglected.

To understand redshifts due to the expanding universe we need to see how length is defined during the expansion.  The metric adopted to define length is the Robertson-Walker metric.  It is the solution to Einstein's field equations under the simplifying assumption of a homogeneous and isotropic universe.  These assumptions are appropriate for this problem because measurements confirm that on scales larger than about 100\,Mpc (1 parsec = 3.26 lighyears = 31 trillion km) the universe is uniformly dense, and therefore homogeneous and isotropic.  The Robertson-Walker metric expresses the observed length $ds$, in terms of the space-time elements in spherical coordinates $(t, r, \theta, \phi)$ as
\begin{equation}
ds^2 = -c^2dt^2 + a^2(t)[dr^2 + S_\kappa^2(r) d\Omega^2].
\label{eqn:metric}
\end{equation}
Here, $c$ is the speed of light, $a(t)$ is a dimensionless scale factor that describes the spatial expansion of the universe, $d\Omega^2 = d\theta^2 + {\rm sin^2}\theta\,d\phi^2$, and $S_\kappa(r)$ accounts for the curvature of space and is given by
\begin{equation}
S_\kappa(r) = \begin{cases}
R_0 \sin(r/R_0) &\mbox{if $\kappa = +1$} \\
r &\mbox{if $\kappa = 0$} \\
R_0 \sinh(r/R_0) &\mbox{if $\kappa = -1$},
\end{cases}
\end{equation}
where $R_0$ is the radius of curvature of the universe and $\kappa$ is the sign of the curvature.  Our universe appears to be flat with $S_\kappa(r) = r$, but the metric can also model universes with positive curvature ($\kappa =+1$) and negative curvature ($\kappa = -1$).  Notice that the flat metric reduces to the familiar spherical coordinates with the addition of the special relativistic term $-c^2 dt^2$, and the expansion scale factor $a(t)$.  It is convenient to define the value of the scale factor to be unity at the present time: $a(t_0) = 1$.  This definition implies that the scale factor was smaller than one in the past.
 
We use light traveling between fixed emission and observation points to measure the distance interval.  Let's set the origin of the coordinate system at the telescope that observes the light.  In this coordinate system a photon travels radially toward the observation point at constant angles ($\theta$ and $\phi$) from the emitting source, giving $d\Omega = 0$.  Furthermore, because light travels along null geodesics defined by  $ds = 0$, we can solve Eq.~\eqref{eqn:metric} to get the distance interval
\begin{equation}
dr = \frac{c\,dt}{a(t)}.  
\label{eq:CDI}
\end{equation}
Hence, the distance interval is determined by the speed of light and the expansion of the universe during transit.

Two important results arise from this equation.  First, if we consider the wavelength of a photon $\lambda = c \Delta t$, where $\Delta t$ is the period of the photon's oscillation, we find that this wavelength is given by $\lambda_e=a(t_e)\Delta r$ at the emission time and $\lambda_{ob}=a(t_{ob})\Delta r$ when it is observed.  The coordinate system itself ($r$, $\theta$, $\phi$) is independent of time, so for observers at fixed coordinates, $\Delta r$ is independent of time.  Using Eq.~\eqref{eq:redshift}, we then find that the scale factor is related to the redshift by
\begin{equation}
\label{eqn:zOfa}
z = \frac{1}{a(t_e)} - 1,
\end{equation}
where we have used our definition that $a(t_0) = 1$ for observations at the present time ($t_{ob}=t_0$).   Thus, by measuring the wavelength shift of an astronomical source and determining the redshift, we can deduce the value of the scale factor at the time the light was emitted.
  
The second important result arising from Eq.~\eqref{eq:CDI} is the calculation of the line-of-sight distance to the source at the time of observation.  This is called the conformal distance $D_c$, and is given by 
\begin{equation}
\label{eqn:dp}
D_c = \int^{D_c}_0 dr = \int^{t_e}_{t_{ob}} \frac{c\,dt}{a(t)}.
\end{equation}
Most of us are familiar with the proper distance.   In special relativity it is defined as the distance between events occurring at the same instant of time ($dt=0$).  The radial proper distance ($d\Omega=0$) is then given by
\begin{equation}
D_p(t) = a(t)\int_0^{D_c} dr = a(t)D_c.
\label{eq:prop_dist}
\end{equation}
Today, the radial proper distance is equal to the conformal distance; in an expanding universe the proper distance was smaller when the light was emitted $D_p(t_e) = a(t_e)D_c < D_c$.

Now that we have a handle on the definitions of distances in the Robertson-Walker metric, we can define the quantities important to astronomers.  To determine a source's intrinsic brightness and physical size we must convert our observations into the object's rest frame.  The observed angle subtended by a galaxy on the sky $\Delta \theta$, is related to the diameter of the galaxy in its rest frame at the time the light was emitted according to
\begin{equation}\label{eqn:galDiameter}
{\rm galaxy\ diameter} = D_A \Delta\theta,
\end{equation}
where $D_A$ is called the angular diameter distance by astronomers.   Equation~\eqref{eqn:metric} shows that this distance is given by
\begin{equation}\label{eqn:da}
D_A = a(t_e) S_\kappa(D_c).
\end{equation}
In a flat universe, the angular diameter distance is just the proper distance to the galaxy at the time the light was emitted.

Similarly, the observed brightness of a source depends on how far away it is.  Experimentally the flux is determined by dividing the energy detected ($E_{ob}$) by the area of the detector's aperture ($A_{\rm det}$) and the exposure time ($\Delta T_{ob}$).  
\begin{equation}
{\rm flux}_{ob} = \frac{E_{ob}}{\Delta T_{ob} A_{det}} 
\end{equation}
To study the emitting source, we have to transform this expression into the rest frame of the emitter.  Because the universe expands while the photons are in transit, the observed exposure time is dilated [$\Delta T_{ob} = \Delta T_e/a(t_e)$] and the photons' energy is lowered due to redshift [$E_{ob} = E_e a(t_e)$].  Also, since photons are emitted by a source in all directions the fraction that make it into a fixed aperture at a distance $r$ is $A_{\rm det} /(4\pi r^2)$.  Here the surface area in the denominator is determined for the detected light at the present time with radius $r = a(t_0) S_\kappa(D_c)$.   With these modifications the observed flux can be expressed as
\begin{equation}
{\rm flux}_{ob} = \frac{E_{ob}}{\Delta T_{ob} A_{det}} = \frac{E_{e} a^2(t_e)}{\Delta T_{e} 4\pi S_\kappa^2(D_c)}
\end{equation}
where we have once again set $a(t_0)$ to unity.
The technical name for the source brightness is luminosity and it is the energy emitted per second ($E_e/\Delta T_e$).  Astronomers relate flux and luminosity with the following relation,
\begin{equation}\label{eqn:flux}
{\rm flux}_{ob} = \frac{ {\rm source\ luminosity} }{4\pi D_L^2},
\end{equation}
where $D_L$ is called the luminosity distance.   The luminosity distance is defined to make the denominator look like a surface area even though there is a lot physics hidden in $D_L$.  $D_L$ is given by
\begin{equation}\label{eqn:dl}
D_L = S_\kappa(D_c) / a(t_e).
\end{equation}

Finally, it is rare to measure a luminosity or flux directly.  Usually we find ourselves taking the logarithm of Eq.~\eqref{eqn:flux} and expressing this in what are called \textit{magnitudes}.  Astronomers define the apparent magnitude by $m = -2.5\log[{\rm flux}/(2.53\times 10^{-8} {\rm watt/m}^2)]$, and the absolute magnitude by $M=-2.5\log[{\rm luminosity}/(78.7 L_{solar})]$.  The absolute magnitude of a source is related to the apparent magnitude by
\begin{equation}
M = m - 5\log\left(\frac{D_L}{10}\right),
\end{equation}
where $D_L$ is given in parsecs.  The second term in this equation is called the distance modulus: $DM = 5 \log(D_L/10)$. Notice that the distance modulus is determined theoretically and can be computed directly from the metric at any emission time.  A prediction of $DM$ exists for every specific cosmological model of $S_\kappa(r)$ and $a(t_e)$.
Direct tests of the expansion have been made by measuring the apparent magnitude for sources with known absolute magnitude and comparing the difference ($m-M$) to the predicted distance modulus ($DM$).  

In summary, from the angular diameter distance and the luminosity distance we can compute the rest-frame size and intrinsic brightness or luminosity of a source emitting radiation.  These distance factors allow us to study the physics of the emitters.  To compute these quantities, we need to determine the conformal distance of the source given by Eq.~\eqref{eqn:dp}.  But to calculate $D_c$ we must first find the scale factor $a(t)$ that describes the expanding universe. This scale factor is found by solving the Friedmann equations.

\section{General Solution to the Friedmann Equations}
The Friedmann equations model a fluid or several fluids composed of different types of particles moving under the influence of gravity in the Robertson-Walker metric. Although the full derivation is beyond the scope of this paper, we will introduce the underlying physics.  

A description of the physics begins by considering conservation of energy in a self-gravitating pool of matter.  The classical picture is that of a small test mass $m$, embedded at a radius $s$, within an expanding sphere of matter.  The kinetic energy of the test mass is $K=\frac{1}{2}m\dot{s}^2$ and the potential energy is $U_g=-GMm/s(t) - U_{\textrm{zero}}$, where $M$ is the mass inside the volume of radius $s$ and $U_{\textrm{zero}}$ is the zero point of the potential energy.  The sum of kinetic and potential energy is a constant:
\begin{equation}
E_{tot} = \frac{1}{2}m\dot{s}^2 + \frac{-GMm}{s(t)}  - U_{\textrm{zero}}.
\end{equation}
To change this into the classical version of first Friedmann equation, we need to make three modifications.  First, we divide by the test mass $m$ and express everything in terms of a potential, e.g. $U=U_{\textrm{zero}}/m$.  Second, we define the value of the zero point so that the total energy is zero at all times.    Third, we use the Robertson-Walker metric to write $\dot{s} = \dot{a} D_c$.  With these changes, we have
\begin{equation}\label{eqn:expand}
0 = \frac{1}{2}D_c^2\dot{a}^2 - \frac{G M}{D_c^2 a^2(t)} - U.
\end{equation}
If we imagine that the Big Bang resulted in a high density fluid [$\rho(t) = M/\frac{4}{3}\pi s^3(t)$] expanding radially due to a large amount of kinetic energy, then the expansion ($\dot{a}$) will slow in time as the scale factor ($a$) increases.  The question then becomes, does the expansion ever actually stop?  The answer to this depends on the value of $U$.  If $U$ is positive, $\dot{a}$ can never be zero.  If $U$ is negative, $\dot{a}$ will become zero when the two right-most terms cancel.  The equation works equally well for expansion ($\dot{a}>0$) and for contraction ($\dot{a}<0$), but because only $\dot{a}^2$ is specified, the sign must be determined by the context.

Equation~\eqref{eqn:expand} can be rearranged into the classical form of the first Friedmann equation 
\begin{equation}\label{eqn:classical}
\left(\frac{\dot{a}}{a}\right)^2 = \frac{8\pi G}{3}\rho(t) + \frac{2U}{D_c^2a^2(t)}.
\end{equation}
Einstein's general theory of relativity leads to a similar equation for energy conservation with two important differences. \cite{peebles, ryden} First, the energy of all particles
($E_i = \sqrt{m_i^2c^4+p_i^2c^2}$)
contributes to the potential energy and not just the mass.  This means massless particles such as photons also contribute to the potential energy.  We therefore replace the mass density $\rho(t)$ with the energy density $\sum\epsilon_i(t)/c^2$, where $i$ is the sum over all particle species. Second, the potential energy depends on the curvature of the universe.  We state without proof that $2U/D_c^2 = -\kappa c^2/R_0^2$.  It is also convenient to define the critical density $\epsilon_c = 3c^2H_0^2/8\pi G$, where $H_0$ is Hubble's constant, defined by the present rate of expansion: $H_0 = \dot{a}({t_0})$.  The result is the fully relativistic Friedmann equation
\begin{equation}\label{eqn:fried}
\left(\frac{\dot{a}}{a}\right)^2= H_0^2\frac{\sum \epsilon_i(t)}{\epsilon_c} - \frac{\kappa c^2}{R_0^2a(t)^2},
\end{equation}
which has the same form as the classical Friedmann equation given in Eq.~\eqref{eqn:classical}.

Before we solve Eq.~\eqref{eqn:fried}, it is interesting to look at different types of fluids and how their energy densities evolve with the scale $a(t)$.  Various known and hypothetical energy densities are shown in Table~\ref{tab:density}.~\cite{adventures}  These are derived by considering the thermodynamic properties of each type of energy density during an adiabatic expansion.  The matter density $mc^2/V$, for example, evolves as $1/a^3(t)$ because the volume ($V$) expands as $a^3(t)$.  The radiation density $hc/\lambda V$, on the other hand, goes as $1/a^4(t)$ because the volume expands as $a^3(t)$ while the wavelength expands as $a(t)$.  Although Table~\ref{tab:density} is organized in increasing powers of $a(t)$, there is no deep significance to this ordering.  In a $\Lambda$CDM cosmology the energy density of the universe has three main components: a cosmological constant [$\Lambda \rightarrow\epsilon_{\Lambda}(t)$], matter [non-relativistic particles, either dark or normal baryonic $\rightarrow\epsilon_{m}(t)$], and radiation [photons and highly-relativistic particles such as low-mass neutrinos $\rightarrow\epsilon_{r}(t)$].  
While other components may exist at a low level, experiments are not yet sensitive to their effects.
By setting $\epsilon_{w,0}$, $\epsilon_{cs,0}$, and $\epsilon_{q,0}$ equal to zero we reduce the general solution to the $\Lambda$CDM case which shows spectacular agreement with observations.

\begin{table}[b]
\begin{center}
\caption{\label{tab:density}Evolution of energy densities considered in cosmology.  It is also common to find these expressed as fractional energy densities, $\epsilon_i(t)/\epsilon_c = \Omega_{i,0}/a^i$.}
\vspace{6 pt}
\begin{tabular}{l c @{\quad} c}
\hline\hline
{\bf Component} & {\bf Index $i$ } & {\bf Energy density $\epsilon_i(t)$}  \\
\hline
Cosmological constant & 0 & $\epsilon_{\Lambda,0}$ \\
Domain walls		& 1 & $\epsilon_{w,0}/a(t)$  \\
Cosmic strings		& 2 & $\epsilon_{cs,0}/a^2(t)$  \\
Matter (non-relativistic) & 3 & $\epsilon_{m,0}/a^3(t)$  \\
Radiation			& 4 & $\epsilon_{r,0}/a^4(t)$  \\
Quintessence		& varies & $\epsilon_{q,0}/a^i(t)$  \\
\hline\hline
\end{tabular}
\end{center}
\end{table}

The first step in the solution is to set the boundary conditions by evaluating the Friedmann equation at the present time $t_0$, for which $a(t_0) = 1$.  As mentioned, the time derivative of the scale factor was first measured by Edwin Hubble in 1929 and is known as the Hubble constant $H_0=\dot{a}({t_0})$.  Finally, we define the fractional energy density $\Omega(t) = \sum \epsilon_i(t)/\epsilon_c$, and note that the fractional energy density is presently $\Omega_0 = \sum \epsilon_i(t_0)/\epsilon_c$.  At the present time, Eq.~(\ref{eqn:fried}) becomes
\begin{equation}
H_0^2 = H_0^2\Omega_0 - \frac{\kappa c^2}{R_0^2},
\end{equation}
which allows us to express the curvature ($\kappa$, $R_0$) in terms of $H_0$ and $\Omega_0$ as
\begin{equation}\label{eqn:curvature}
\frac{\kappa c^2}{R_0^2} = H_0^2(\Omega_0-1).
\end{equation}
In this form we see that if the total density, $\sum \epsilon_i(t_0)$, is greater than the critical density $\epsilon_c$, then the curvature is positive and $R_0 = \sqrt{c^2/H_0^2(\Omega_0-1)}$.  If  the total density is less than the critical density, then the curvature is negative and $R_0 = \sqrt{c^2/H_0^2(1-\Omega_0)}$.  These constants are not time dependent and therefore the curvature will be independent of time.  

Rewriting the Friedmann equation in terms of constant fractional densities allows us to show the dependence on the scale factor explicitly
\begin{equation}\label{eqn:ode}
\left(\frac{\dot{a}}{a}\right)^2= H_0^2\sum \frac{\Omega_{i,0}}{a^{i}} + \frac{H_0^2(1 - \Omega_0)}{a^2}.
\end{equation}
The solution to this first-order separable differential equation is
\begin{equation}\label{eqn:lookback}
H_0 \int_{t_0}^{t_e} dt = \int_1^a \frac{da^\prime}{\sqrt{\sum \Omega_{i,0} /a^{\prime(i-2)} + (1 - \Omega_0)}}.
\end{equation}
where we integrate from $a(t_0) = 1$ to a time in the past, $a(t_e)$.  If we set $t_0=0$, then $t_e$ is called the look-back time and is the time measured backwards from today.  In the limit $a\rightarrow0$, $t_e$ equals the age of the universe.  Another choice is to set $t_e$ equal to zero, in which case $t_0$ becomes the age of the universe.  

Unfortunately, the integrand on the right-hand side is not easily integrated so numerical methods are typically required.  In fact, analytic integration is only possible for simplified universes in which some components are ignored.  Such solutions can be useful for times when one component dominates the expansion.  Two special cases are particularly useful for debugging the numerical routine.  The first is a flat universe containing only matter ($\Omega_0 = \Omega_{m,0} = 1$), where the analytic solution is $a(t_e) = (3H_0\, t_e/2)^{2/3}$.  The second is a flat universe containing matter and lambda ($\Omega_0 = 1 = \Omega_{m,0} + \Omega_{\Lambda,0}$), where the analytic solution is $a(t_e) = (\Omega_{m,0}/\Omega_{\Lambda,0})^{1/3} \sinh^{2/3}(3\sqrt{\Omega_{\Lambda,0}}\,H_0\,t_e/2)$, for $0<\Omega_{m,0}<1$.

It is common to use a change of variables to write the solution in terms of the conformal distance.~\cite{hogg}  Using $dr = c\,dt/a(t)$ from Eq.~\eqref{eqn:dp} leads to an alternative form of Eq.~\eqref{eqn:lookback} given by
\begin{equation}\label{eqn:confDist}
\frac{H_0}{c} \int_{0}^{D_c} dr = \int_1^a \frac{da^{\prime}}{\sqrt{\sum \Omega_{i,0} /a^{\prime(i-4)}+ (1 - \Omega_0)a^{\prime2}}}.
\end{equation}
In this form, the conformal distance is computed directly instead of the look-back time.  In addition, it is  common to find Eqs.~\eqref{eqn:lookback} and \eqref{eqn:confDist} in the literature expressed as an integral over redshift $z$, where Eq.~\eqref{eqn:zOfa} is used to change variables from scale factor to redshift.

\section{Numerical Solution}
Implicit in the solution to Eqs.~\eqref{eqn:lookback} and \eqref{eqn:confDist} is the judicious choice of integration limits.  Because the integrand is undefined at the start of the universe, it is important to integrate from a time after the Big Bang ($t_e$) until today ($t_0$).  Without extending the model, it is also a good idea to avoid epochs prior to $10^{-32}$ seconds when inflation and quantum gravity play an important role.  

We begin by integrating both sides of Eq.~\eqref{eqn:lookback} numerically to find the scale factor as a function of time.  Although this may seem simple, because we are interested in finding the scale factor as a function of time, we must integrate many times.  We explore the full expansion history by choosing a series of integration limits from today's value ($a=1$) back to a time when the scale factor was very small, say $a=10^{-6}$.  Unfortunately, the calculation naturally produces $t(a)$ instead of $a(t)$. Because the equation cannot be inverted analytically, we limit our solutions to unique single-valued functions and just swap the $a$-axis with the $t$-axis to get $a(t)$.  As an alternative, because Eq.~\eqref{eqn:ode} is an ordinary differential equation (ODE), many computing languages include numerical solvers that will calculate $a(t)$ directly.  Although usually somewhat slower, this method has the advantage of working even when the function is not single valued.  

Direct integration of Eq.~\eqref{eqn:lookback} can be accomplished in almost any computing language using Simpson's Method or a faster Romberg algorithm~\cite{numericalRecipes}  and the numerical integration routines in MATLAB and MATHEMATICA are based on these algorithms.  Many students taking astronomy classes, however, may not be proficient in these languages, so an alternative is to use a trapezoid method in EXCEL.~\cite{excelIntegration}  In fact, because we integrate repeatedly, the trapezoid method avoids evaluating the integrand multiple times.  First, a column or array is created for the scale factors stepping backward from $\log_{10}(a)=0$ to $\log_{10}(a)=-6$.  More columns are needed to compute the area of each trapezoid formed by the step size $\Delta a$, and the average of the integrand computed at $a_i$ and $a_{i-1}$.  The solution is then computed as a running sum in the final column.  This column is then relabeled $H_0(t_e-t_o)$, the left-hand-side of Eq.~\eqref{eqn:lookback}.  Another column for the age can be computed by setting $t_e = 0$ years and solving for $t_0$.

\begin{table}[b]
\begin{center}
\caption{\label{tab:const}For precise computations, constants and conversion factors
with six significant figures must be used. }
\vspace{6 pt}
\begin{tabular}{c@{\quad} c@{\quad} c}
\hline\hline
{\bf c  (km/s)} & {\bf sec/year } & {\bf Mpc/km }  \\
\hline
$2.99792\times10^5$ & $3.15581\times10^7$ & $3.24078\times10^{-20}$  \\
\hline\hline
\end{tabular}
\end{center}
\end{table}

\begin{figure}[t]
\includegraphics[scale=0.85]{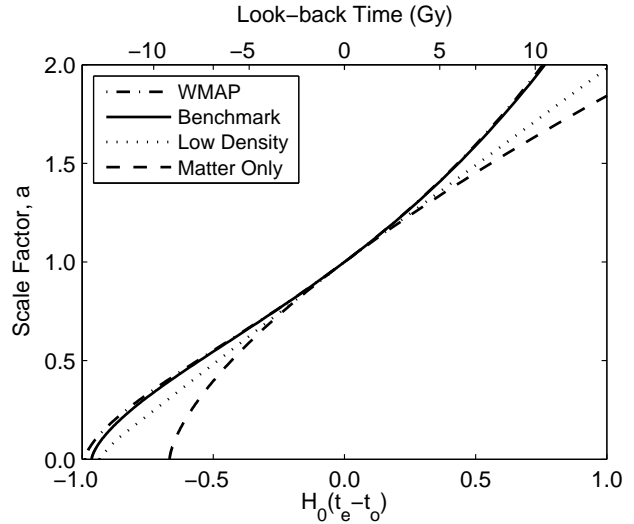}
\caption{\label{fig:scale} Scale factor $a(t)$, for the parameter sets described in Table \ref{tab:params}.  Notice that $a(t)$ and $H_0(t_e - t_0)$ are unitless.  The scale factor has a value of 1 today.  For the parameter sets shown, it is smaller in the past and larger in the future.}
\end{figure}

It's a good idea to check the numerical integration using the analytical solutions for a few special cases as discussed above.  Plotting the scale factor as a function of the look-back time, age, or redshift shows the expansion of the universe visually.  As an example, Fig.~\ref{fig:scale} shows $a(t_e)$ for a variety of cosmologies.   The slope and curvature of this plot can be interpreted as the speed and acceleration of the spatial expansion of the universe.  The numerical derivatives ($\dot{a}$ and $\ddot{a}$) are also easy to compute and visualize graphically.  The figures in this paper were created using MATLAB but they are easily created in any computing language.

\begin{table*}[t]
\begin{center}
\caption{\label{tab:params}Four sample parameter sets of interest.  The WMAP parameters 
are taken from the seven-year release~\cite{wmap} plus radiation.~\cite{pdg,komatsu,larson}  The Benchmark parameters are a simplified one-significant figure set circa 2003 that are in Ryden's textbook.~\cite{ryden}  The Low Density parameters include only radiation and baryonic densities.  The Matter Only parameters set the matter density to the critical density.}
\vspace{6 pt}
\begin{tabular}{l c @{\quad} l @{\quad} l @{\quad} l @{\quad} l}
\hline\hline
& $H_0$ (km/(s Mpc))  & $\Omega_{m,0}$ & $\Omega_{r,0}$  & $\Omega_{\Lambda,0}$ & $\Omega_0$\\
\hline 
{\bf WMAP 7-year } & 70.4 & 0.2722 & 8.42E-5 &  0.728 & 1.000000\\
\hline
{\bf Benchmark }     & 70   & 0.3       & 8.4E-5        & 0.7 & 1.000084\\
\hline
{\bf Low Density  }  & 70   & 0.05    & 8.4E-5        & 0 & 0.050084\\
\hline
{\bf Matter Only  }   & 70   & 1.0       & 0                 & 0 & 1.0\\
\hline\hline
\end{tabular}
\end{center}
\end{table*}

\begin{figure}[t]
\includegraphics[scale=0.8]{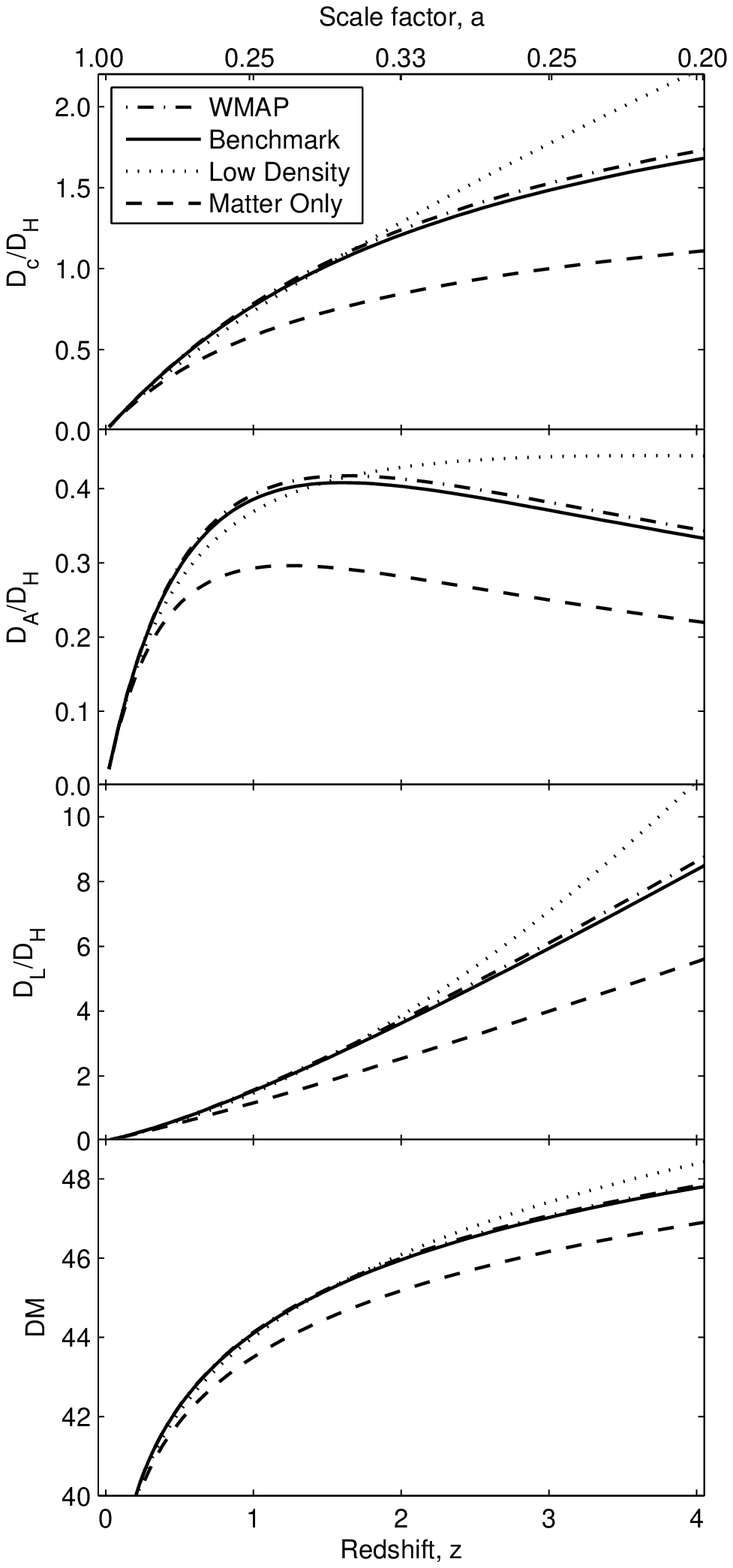}
\caption{\label{fig:distanceFactors} Conformal distance (top), angular diameter distance (2nd from top), luminosity distance (2nd from bottom), and distance modulus (bottom).  The top three distance factors are divided by the Hubble distance $D_H = c/H_0 \sim 1600$~Mpc. }
\end{figure}

To compute four significant figures precisely, it is important to use a small step size and physical constants with at least six significant figures of precision.  Table~\ref{tab:const} shows the constants and conversion factors used in our calculations.  To sample smoothly at all time scales, steps in the logarithm of $a$ are preferable to linear steps.  Comparisons with analytical solutions show that a step size of $\Delta\log_{10}(a) = 0.01$ results in at least four significant figures of precision.  Table~\ref{tab:params} shows the cosmological parameters for the test cases included in Fig.~\ref{fig:scale}.  The density parameters are expressed as fractions of the critical density:  $\Omega_{m,0}=\epsilon_{m,0}/\epsilon_c$ is the fractional non-relativistic matter density that includes normal and dark matter, $\Omega_{r,0}=\epsilon_{r,0}/\epsilon_c$ is the fractional relativistic energy density that includes photons and neutrinos, and $\Omega_{\Lambda,0}=\epsilon_{\Lambda,0}/\epsilon_c$ is the fractional energy density attributed to the cosmological constant.  Notice that we have explored only a subset of the possible parameters.  Projects involving cosmic strings or quintessence could be easily implemented by including more terms in Eq.~\eqref{eqn:lookback}.  

Here we focus on the recent 7-year release from the Wilkinson Microwave Anisotropy Probe (WMAP) collaboration.~\cite{wmap}  These density parameters are obtained by fitting a combination of data from WMAP measurements, the Sloan Digital Sky Survey measurements of baryon acoustic oscillations,~\cite{sdss} and the Hubble Space Telescope measurements of $H_0$.~\cite{hst}  These fits yield the best observationally-based determinations of $H_0$, $\Omega_{m,0}$, and $\Omega_{\Lambda,0}$.  The best estimate for the radiation density still comes from the Cosmic Observation Background Explorer (COBE) collaboration,~\cite{cobe, pdg} and the neutrino energy density is included according to the theoretical expectation for the effective number of neutrino species: $N_\textrm{eff} = 3.04$.~\cite{komatsu, larson}  Because the measured curvature is consistent with zero and theoretically this is difficult to explain unless the universe is indeed flat, the WMAP team set $\Omega_0 = 1$.

Table~\ref{tab:params} also shows the Benchmark parameters suggested in Ryden's textbook.~\cite{ryden}  These are one-significant figure estimates suitable for homework calculations.  The Low Density parameters are chosen because most of us want to know what our universe would look like with just the ``normal" stuff we have studied in the laboratory.  Because these parameters are so small, this Low Density universe is highly curved.  A similar motivation leads us to explore the Matter Only parameters containing enough matter that this universe is flat.

The goal of this calculation is to compute ages and distance factors. So far we have only computed the age of the universe.  We also want to compute the various distance factors.  Although it is convenient to compute the conformal distance by integrating Eq.~\eqref{eqn:confDist}, having students compute the integral in Eq.~\eqref{eqn:dp} instead reinforces the idea that the conformal distance is found by integrating over time as the universe expands. To compute the angular-diameter and luminosity distances, we need to know the radius of curvature of the universe $R_0$, and the sign of the curvature $\kappa$, computed from Eq.~\eqref{eqn:curvature}.  Then it is simple to find $D_A$ and $D_L$ from Eqs.~\eqref{eqn:da} and \eqref{eqn:dl}. Finally we compute the distance modulus $DM = 5 \log(D_L/10)$ remembering to convert $D_L$ into units of parsecs.  Figure~\ref{fig:distanceFactors} shows the conformal distance, luminosity, and angular diameter distances as well as the distance modulus for a variety of cosmological models.   The calculation takes less than 0.5 seconds in EXCEL or MATLAB.

A numerical issue arises when we attempt to find distance factors and ages based on redshift.  The tabulated array of $\log_{10}(a)$ can easily be used to compute an array of redshifts but it is unlikely that the exact redshift has been computed.  With a reasonably small step size in $\log_{10}(a)$, simple linear interpolation gives accurate values because the expansion of the universe is smooth and well behaved.

An example of a class project based on this technique is available online~\cite{me, supplement} and in the supplementary materials.   The project is written in four parts and assumes that students are familiar with the theory presentation in Ryden's textbook. The first part uses the Matter Only test case $H_0(t_e-t_o) = \frac{2}{3}(a^{3/2}-1)$, to compute the age and distance factors $D_C$, $D_A$, $D_L$, and $DM$ as a function of the scale factor or redshift.  General instructions are given for advanced students who want to choose their own computing language.  Detailed instructions are given in EXCEL for less advanced students.  A screen shot of the solution in EXCEL helps students debug their calculations.  Students report the age and horizon distance, recreate the Matter Only curves in Figs.~\ref{fig:scale} and~\ref{fig:distanceFactors}, and interpret the results.    

In the second part of the project, we replace the Matter Only test case with the general solution to Eq.~\eqref{eqn:lookback}, which involves another numerical integration.  After debugging the solution by comparing to the analytical solution of part 1, students compute Table~\ref{tab:compare} to compare different cosmologies.  In the final parts of the project, students explore the applications described in the next section.  An important aspect of these applications is that students are directed to original sources online to find the latest results for themselves.

Advanced students are able to complete the project in about six hours.  Less advanced students require instruction on creating parameters and using them appropriately in equations in EXCEL.  Submitting the first part of the project well ahead of the others allows students time to recover from many of the bugs they may encounter.  In the later parts of the project it's important that the bugs are gone so that they can focus on the physical implications of the integration.

\section{Observational Applications}
Students can use these results to compute rest-frame properties such as luminosity and the physical dimensions of sources. To help students appreciate the physical implications of their calculations, they must apply them to four different problems: 1) find the distance and size of a high redshift object; 2) find the age and horizon of the WMAP seven-year universe; 3) compare the WMAP seven-year universe to RydenÕ's Benchmark universe; 4) examine why the low-density model is not acceptable compared to the WMAP model.

\subsection{ Galaxy Zoo: Hubble (Zooniverse)}
\begin{figure*}[t]
\includegraphics[scale=0.55]{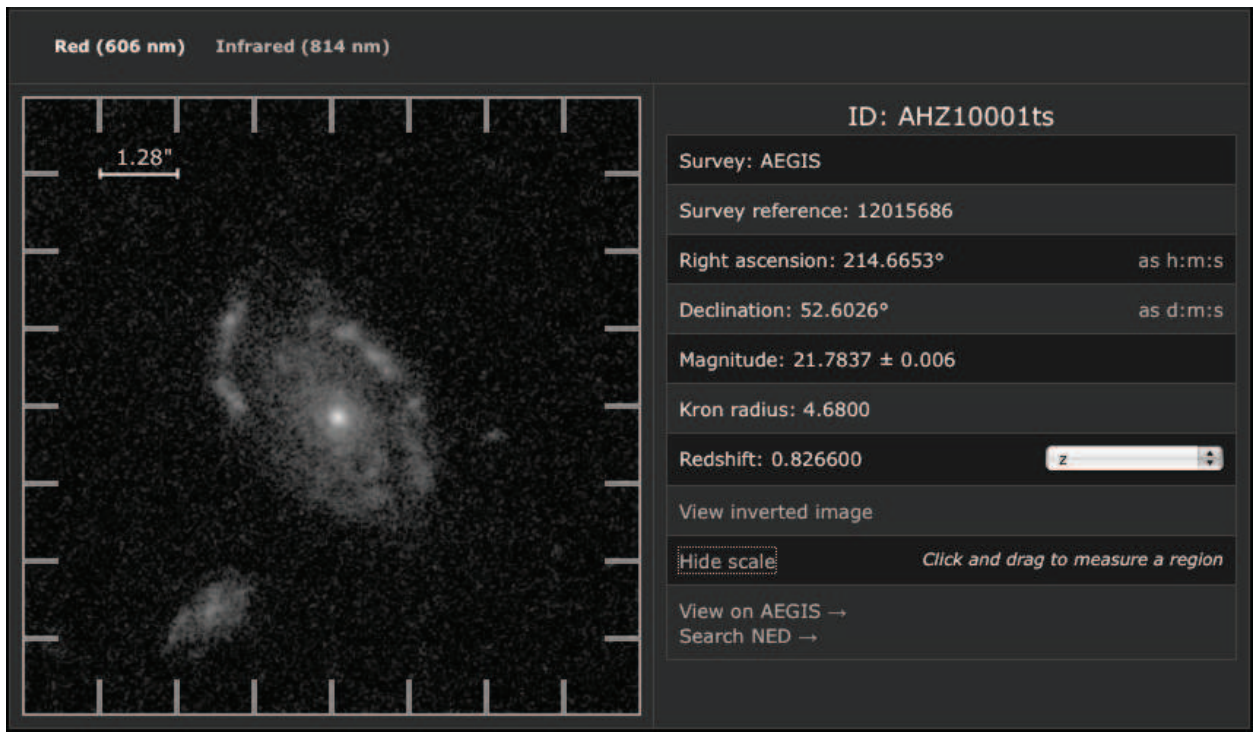}
\caption{\label{fig:galaxyZoo} Image courtesy of {\it Galaxy Zoo: Hubble} project.~\cite{galaxyZoo}  This galaxy is at a redshift of 0.8266 corresponding to an emission time of 7.01 billion years ago.  Its angular size in the long dimension is about $5.12''$, which corresponds to 39\,kpc.  If the neighbor galaxy is at the same redshift, it is located 29\,kpc away from the larger galaxy and has a diameter of 18\,kpc. }
\end{figure*}

{\it Galaxy Zoo: Hubble}~\cite{galaxyZoo} is an excellent website to introduce students to galaxy classification. It doesn't take long to find a few complicated objects that make one ponder about the sizes of the features in the images.  In {\it Galaxy Zoo: Hubble} students can save interesting galaxies into an album where more detailed information can be displayed.  An example of such a galaxy is shown in Fig.~\ref{fig:galaxyZoo}.  

From the given redshift of $z = 0.8266$, our numerical calculation tells us that the angular diameter distance to the central galaxy is $D_{\rm A} = 1580$\,Mpc and that the look-back time is $t_e = 7.01$ billion years.  We therefore learn that the light observed by the Hubble Space Telescope in this image was emitted when the universe was slightly less than half its current age.  Using the angular scale on the image we can estimate the angular extent of the larger galaxy along its major axis to be $\Delta \theta = 5.12''$.  Using Eq.~\eqref{eqn:galDiameter} we then find that the galaxy's diameter is 39\,kpc.  

It is interesting to compare the dimensions of these distant objects to the dimensions of galaxies in our local group of galaxies.  The size of the larger galaxy is similar to the size of the Milky Way galaxy.  If the small neighboring galaxy is at the same redshift and not a chance alignment, it measures 18\,kpc along its major axis which is about four times larger than the Large Magellanic Cloud.  The two galaxies are separated by about 29\,kpc, which is similar to the distance to our closest neighboring galaxies.  These comparisons show that gravity in the past produced structures with similar sizes and separations to those of galaxies in our vicinity today.

\subsection{Age and Horizon of the WMAP Seven-year Universe}
Even after a course in cosmology it is likely to remain a bit of a mystery how professional cosmologists calculate the precise age and size of our visible universe.  Here we accurately reproduce the latest state-of-the-art results presented by the WMAP team~\cite{wmap} to show all steps of the calculation clearly.  Table~\ref{tab:compare} compares the published values to our calculation  performed with the WMAP seven-year parameters listed in Table~\ref{tab:params}.  The published age of the universe is $13.75\pm0.11$ billion years and our computation reproduces this result.  It is worth mentioning that the experimental result has two significant figures and that, within the experimental uncertainty, this age may change in the coming years.  However, as new, more precise density parameters become available, we will continue to be able to compute the age with a precision of four significant figures.

\begin{table*}[t]
\begin{center}
\caption{\label{tab:compare}The published WMAP~\cite{wmap} ages and distance factors compared to our computed values.  Notice that our calculation reproduces at least four significant figures, which is more numerical precision than the experimental uncertainty.   It is also interesting that the Benchmark universe is accurate to about 3\%, which although close, is not in agreement with the published WMAP values.}
\vspace{6 pt}
\begin{tabular}{l @{\quad} c @{\quad} c @{\quad} c @{\quad} c @{\quad} c }
\hline\hline
& Age & Horizon & Age at & $D_c$ at  & $D_c$ at \\
& (billion yrs) & distance &  decoupling &    decoupling  & Matter-Radiation \\
&                      &  (Mpc)     &  (yrs)              &  (Mpc)             & Equality (Mpc)       \\
\hline
{\bf PUBLISHED:} & & & & & \\
{\bf WMAP 7-year}&  $13.75\pm0.11$  &   & $377,730^{+3205}_{-3200}$  &  $14,073^{+129}_{-130}$ & $14,238^{+128}_{-129}$  \\
\hline
{\bf THIS WORK: }& & & & & \\
{\bf WMAP 7-year} & 13.75 & 14,357 & 377,710 & 14,073 & 14,238 \\
\hline
{\bf Benchmark }     &  13.46  & 13,897 & 368,521 & 13,618  &  13,780 \\
\hline
{\bf Low Density  }  & 13.04  & 17,802 & 548,438 & 17,421  & 17,664 \\
\hline
{\bf Matter Only  }   & 9.31    &  8,558 & 258,085 & 8,307  & 8,416 \\
\hline\hline
\end{tabular}
\end{center}
\end{table*}

The size of the visible universe is characterized by the horizon distance.  
The horizon distance is the conformal distance determined for light emitted at the time of the Big Bang.  Any observation of our universe necessarily occurs within a sphere of radius $D_C(t_{\rm BigBang})$.  We calculate the horizon distance to be 14\,357 Mpc for the WMAP universe.

Radiation emitted at the Big Bang was reabsorbed many times in the dense early universe and does not reach our telescopes today.  As the universe expanded it cooled and eventually reached the temperature at which free electrons and protons combine to form neutral hydrogen, an event called recombination.  As recombination was occurring, the mean-free-path of photons was lengthening dramatically because the probability that a photon scatters from a neutral hydrogen atom is much smaller than the probability that it scatters from a charged electron.  Shortly after recombination, therefore, the universe became transparent and the radiation and matter components of the universe decoupled. The cosmic microwave background light that we measure today are the photons that last scattered at about the time of decoupling.  The redshift of decoupling published by the WMAP team~\cite{wmap} is $z=1090.89\pm^{0.68}_{0.69}$.  We interpolate using this redshift and find that the conformal distance to decoupling is $D_C= 14\,073$\,Mpc and that decoupling occurred just 377\,710 years after the big bang.  As noted earlier, radiation energy density scales as $1/a^4(t) = (1+z)^4$. Because the radiation density is proportional to $T^4$, it follows that
\begin{equation}
T_e = T_{ob} (1+z).
\end{equation}
Setting $z = 1\,090.9$ and $T_{ob}= 2.7255$\,K we find that the decoupling occurred when the temperature of the universe was  $T_e = 2976.0$\,K.

Another cosmological marker is the time when the matter density surpassed the radiation density. 
In the early universe, radiation dominated the energy density.  Because the radiation energy density decreases as $1/a^4(t) = (1+z)^4$, which is faster than the $1/a^3(t) = (1+z)^3$ with which the matter energy density decreases, there is a redshift at which the two are equal
\begin{equation}
\Omega_{r,0} (1+z)^4 = \Omega_{m,0} (1+z)^3.
\end{equation}
Solving for $z$ and using the WMAP values of $\Omega_{\rm r,0}$ and
$\Omega_{\rm m,0}$ gives
\begin{equation}
 z = \frac{\Omega_{\rm m,0}}{\Omega_{\rm r,0}}-1 = \frac{.2722}{8.42 \times
10^{-5}} -1 = 3\,232.
\end{equation}
The redshift of matter-radiation equality is published in the WMAP paper~\cite{wmap} as $z=3\,232\pm87$.   We again interpolate to find that the conformal distance to this event is $D_{\rm C}= 14\,238$\,Mpc.

These calculations reproduce four or more significant figures of precision, which is better than the measurement uncertainties. This is not to imply that the measurement uncertainties produce the numerical error, but rather that our numerical precision and accuracy are sufficient for the time being. This exercise shows how professional cosmologists compute these factors.  It also gives us confidence that the calculation is accurate at all observable time and distance scales.

\subsection{Comparison of the WMAP Seven-year Universe to the Benchmark Universe}
Students using Ryden's textbook will be familiar with the Benchmark universe parameters.  These parameters are conveniently simple with just one significant figure, and nice to use in a classroom setting.  It is interesting to compute the difference between the Benchmark Universe (circa 2003) and the best available measurements.  Table~\ref{tab:compare} shows that ages and horizon distances differ from the WMAP seven-year release by about 3\%.  From this difference it is easy to argue that the basic sequence of cosmological events has not changed since 2003 even as better datasets became available.  The results presented in the Ryden textbook are therefore close enough to the most accurate results to be of pedagogical value.  Students generally think 3\% is very good agreement.  The 3\% differences, however, are more than $3\sigma$ which makes for a good discussion of precision versus accuracy.  Although pretty close, the Benchmark model is not a good fit to the data.  This is a rare opportunity for an undergraduate student to witness science as it is progressing. 

\subsection{Doesn't the Low Density universe do a pretty good job?}
Finally, the indirect detection of dark matter and dark energy has generated a lot of enthusiasm in observational cosmology.  This excitement has even made it into the popular media. What is surprising is that a simple Low Density universe made up of just 5\% baryonic matter, a little radiation, and the curvature that would result from general relativity is pretty similar to the WMAP and Benchmark universes out to a redshift of about 1.5 (see Fig.~\ref{fig:distanceFactors}).   The Low Density universe contains only baryonic matter and radiation, the stuff we have experimented with in our laboratories, and none of the exciting new stuff.

\begin{figure}[t]
\includegraphics[scale=1.0]{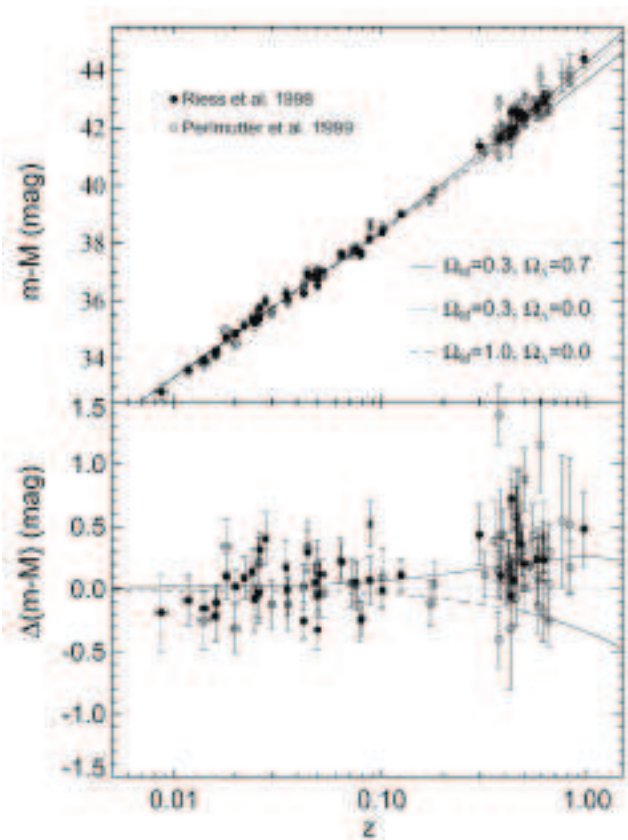}
\caption{\label{fig:snReproduction} Distance Modulus $DM=m-M$, as a function of redshift.~\cite{ryden, riess, perlmutter}.   The bottom panel shows the residuals, \emph{i.e.} the differences between the observed distance moduli and those computed for a universe with $\Omega_{\rm M} = 0.3$ and  $\Omega_\Lambda=0.0$.  This figure is  Reprinted by permission of Pearson Education, Inc., Upper Saddle River, New Jersey.}
\end{figure}

\begin{figure}[t]
\includegraphics[scale=0.8]{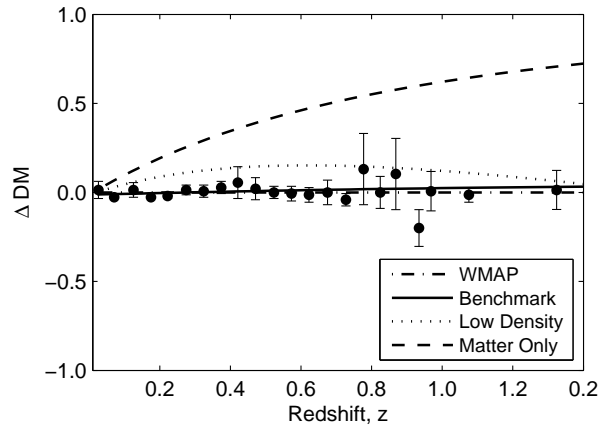}
\caption{\label{fig:snResiduals} Union-2 Compilation~\cite{union2} distance-modulus residuals compared to our test cosmologies.  The Benchmark and WMAP cosmologies are both in agreement with the measured distributions.  The Low Density and Matter Only universes are ruled out in the redshift range from $0.2 <z< 0.8$.}
\end{figure}

At redshifts below 1.5, Type IA supernovae have been used to measure the distance modulus vs. redshift.  The graph of the distance modulus vs. \!$z$ is called the Hubble Plot.  In the bottom panel of Fig.~\ref{fig:distanceFactors} we show Hubble plots computed for four different universes (\emph{i.e.} for four different sets of parameters).  For comparison, we have reproduced the Riess~\cite{riess} and Perlmutter~\cite{perlmutter} Hubble Plots in Fig.~\ref{fig:snReproduction}.  These data come from the papers for which Perlmutter, Schmidt, and Riess won the 2011 Nobel Prize.~\cite{nobel} The solid line in the plot is the Benchmark universe. The Low Density universe falls about 0.1 magnitudes higher than the Benchmark universe at redshifts above 0.3.  The scatter in the data at these redshifts is generally within about 0.2 magnitudes.  When these data were published, the central value favored a cosmological constant.  It is interesting to note, however, that the Low Density universe which lies about 0.1 magnitudes above the solid line was still a $\sim3\sigma$ possibility.  

Today,  the Union-2 SN Compilation~\cite{union2} contains a much more extensive data set.  We have reproduced the more extensive data set in Fig.~\ref{fig:snResiduals} and compared it to the WMAP universe.  This comparison shows that the Low Density and Matter Only universes are now excluded by the data with high probability.  Excluding these alternative universes is an important exercise for many students.  

\section{Conclusions}
In the early sections of this paper we presented a theoretical model of the expanding universe at the undergraduate level.  We also connected the theoretical aspects of the model to astronomical quantities of sources such as redshift, rest-frame size, and brightness.

To accurately calculate the age and various distances in our universe it is important to have consistency between the theoretical presentation and the experimentally determined density factors.  Although specialists in the field routinely compute these age and distance factors, the WMAP seven-year density parameters in Table~\ref{tab:params} are not yet in the literature in this form.  In the literature, $\Omega_{m,0}$ is divided into separate baryonic (normal) and dark matter components.  The simple addition of these leads to round-off errors.  Instead $\Omega_{m,0}$ has to be computed from the more fundamental measurement of $\Omega_{m,0}(H_0/100)^2$. Also, since the $\Omega_{r,0}$ is not measured directly by WMAP, it is not presented with the other parameters.  Only by digging into the details of the papers do we find that WMAP used the value presented here.  Most astronomers are familiar with the theoretical presentation given in this paper and use it to compute high redshift distances and ages.  We therefore believe it is useful to communicate the central value of the WMAP seven-year density parameters in this form.

We are enthusiastic about the educational value of numerically computing accurate age and distance factors in cosmology.  The calculations presented here are accurate back to just moments after the Big Bang.  In this paper, we have described four projects that can be computed in any language including EXCEL to build our understanding of the Standard Model of Cosmology.   These calculations remove much of the mystery in how the age of the universe is determined by professional cosmologists.  Comparison with published values yields four significant figures of precision which is sufficient for the time being.  We also show that galaxies at high redshift have similar gravitational characteristics to those in our nearby vicinity.  This is presumably due to the nature of gravitational attraction remaining constant over the history of the universe.  Finally, we show that the Low Density and Matter Only universes are clearly inconsistent with the Hubble Plot constructed with the largest compilation of Type IA supernovae to date.    Each of these projects gives hands-on experience bringing observations and theory together to build a better understanding of reality.

\begin{acknowledgments}
J. Christiansen would like to thank E.V. Linder for introducing her to these numerical techniques and for many useful discussions.  She also thanks C. Spitzer, E. Albin, K.A. James, and T.K. Fletcher for demonstrating that undergraduates can do these calculations in a research setting.  Finally, she would like to thank her Winter 2011 ASTR-326 class for their enthusiasm and encouragement of this project.  This work was supported by NSF Grant No. PHY-0758085 and the Cal Poly College of Science and Math.
\end{acknowledgments}

\end{document}